\documentstyle[aps, preprint]{revtex}

\begin{document}

\preprint{submitted to JOSA B}

\tighten

\title{Spatial holeburning effects in the amplified spontaneous
emission spectra of the non-lasing supermode in semiconductor laser arrays}

\author{Holger F. Hofmann$^{\ast}$ and Ortwin Hess\thanks{permanent address: 
Institute of Technical Physics, DLR, Pfaffenwaldring 38--40, D--70569 Stuttgart, Germany}}
\address{Edward L. Ginzton Laboratory, Stanford University, Stanford, CA 94305, USA.}

\date{\today}

\maketitle

\begin{abstract}
The amplified spontaneous emission spectrum of the light field in the 
non-lasing supermode of two coupled semiconductor lasers is analyzed using 
linearized Langevin equations. 
It is shown that the interference between the laser mode and
the fluctuating light field in the non-lasing mode causes spatial holeburning.
This effect introduces a phase sensitive coupling between the laser 
field and the fluctuations of the non-lasing mode. For high laser fields, 
this coupling splits the spectrum of the non-lasing mode into a triplet consisting of
two  relaxation oscillation sidebands which are in phase with the laser light and a 
center line at the lasing frequency with a phase shift of $\pm \pi/2$ relative 
to the laser light.
As the laser intensity is increased close to threshold, the spectrum shows a 
continuous transition from the single amplified spontaneous emission line at the 
frequency of the non-lasing mode to the triplet structure.  An analytical expression for
this transition is derived and typical features are discussed. 
\end{abstract}


\section{Introduction}
The carrier density dynamics of semiconductor lasers couple not only to 
the lasing modes, but also modify the amplified spontaneous emission 
into the non-lasing modes. It is therefore possible to gain interesting 
insights into holeburning effects by studying the light field fluctuations 
in the non-lasing modes \cite{Hof97,Jan97,Lem97,Hof98}. 
The most simple 
system which can exhibit holeburning
effects is a laser array in which each of the individual lasers can be
considered to operate in a single mode which is coupled to the local carrier
density of that individual laser. A rate equation description for this
type of device was developed by Winful and Wang \cite{Win88,Wan88} and has been 
discussed in a more general context by the authors \cite{Hof98}. In the following,
this model will be referred to as the split density model since the characteristic 
feature of the model is the separation of the carrier density into distinct density 
pools associated with the individual lasers in the array. Originally,
the split density model was intended to describe instabilities  in semiconductor laser
arrays. Indeed, the model for a two laser array  exhibits a region of instability with a
limit cycle \cite{Win88}.
However, the model can also describe the stable anti-phase
locking between neighboring lasers in an array. This type of anti-phase 
locking is commonly observed in vertical cavity surface emitting laser arrays
\cite{Ore92,Mor92,Mor93,Cat96}.
In particular, the  presence of carrier diffusion leading to a stochastical exchange of 
carriers between the lasers was found to enhance the stability of the
anti-phase locking considerably \cite{Hof98}. 
In the case of stable anti-phase locked operation, the special features
of the model are still present in the dynamics of the quantum fluctuations
causing amplified spontaneous emission in the non-lasing 
modes of the array. Specifically, the spatial holeburning
represented by the carrier density difference between individual 
lasers is part of the fluctuation dynamics, modifying the amplified 
spontaneous emission spectrum in the non-lasing modes
according to this coupling between the light field and the carrier 
density dynamics \cite{Hof98a}.  

Experimentally, the symmetric non-lasing mode and the anti-symmetric lasing 
mode can be distinguished in the far field pattern. 
In particular, the spectrum of the non-lasing mode can be obtained
in the center of the double-lobed far field where the anti-symmetric 
lasing mode vanishes
due to destructive interference. At this very point, the non-lasing 
symmetric mode has an intensity maximum as a result of constructive 
interference. Assuming the lasing mode to be much stronger than the non-lasing
mode, it should be possible to locate the center of the far field at the
point where between the two maxima at the edges the total intensity is at 
its (local) minimum.
If the far-field is observed off-center, a superposition of the
laser spectrum and the spectrum of the non-lasing mode should be
obtained. In this case it should be possible to distinguish the narrow 
laser line from the broad spectrum of the non-lasing mode. 
Spectral features of the fluctuations in the lasing mode such as 
relaxation oscillation sidebands should not be significant until the 
lasing mode intensity is much higher than the intensity of
the non-lasing mode. Interference between the laser line and the fluctuations
in the non-lasing mode are not to be expected since constructive and 
destructive interference are equally likely and consequently average out.

In this paper we derive a general expression for the spectrum of the 
non-lasing mode in a two laser array.
In section \ref{sec:SDM}, we introduce the split density model for two
lasers operating in the antisymmetric supermode. The Langevin equation
for the light field in the non-lasing mode is derived and the effects
of spatial holeburning on the fluctuation dynamics are discussed.
In section \ref{sec:spectrum}, the Langevin equation is solved and the
analytic expression of the spectrum is derived. General features of the
spectrum are discussed.
In section \ref{sec:cases}, the spectra of four typical cases are presented,
illustrating the different spatial holeburning features which may arise in
different types of semiconductor array devices.

\section{Split density model of a two laser array}
\label{sec:SDM}
\subsection{Rate equations}
Generally a semiconductor laser array can exhibit a large range of
dynamical effects involving far more than just two light field modes and two 
spatially separated carrier
systems. However, the region of stable anti-phase locked operation 
relevant for at least some of the experimental realizations 
\cite{Ore92,Mor92,Mor93,Cat96}
can be described realistically by a split density model limited to only 
two light field modes and two carrier
densities. The consistency of such a description with
a more detailed model based on partial differential equations has 
been established in \cite{Mun97}. We will use a version of the model which 
includes both the possibility of carrier diffusion and of a difference in
the cavity loss rate between the symmetric and the antisymmetric supermodes.
The rate equations read 
\begin{mathletters}
\begin{eqnarray}  
\frac{d}{dt} E_1 &=& 
 \frac{\mbox{w}}{2}   N_1  (1-i\alpha)E_1 -(\bar{\kappa}+i\bar{\omega})E_1  
-(\frac{s}{2}-i\frac{\Omega}{2})E_2 
\\  
\frac{d}{dt} N_1 &=& \frac{\mu}{2} - \gamma N_1 - \Gamma (N_1 - N_2) 
- 2\mbox{w} E_1^*E_1 N_1 \\
\frac{d}{dt} E_2 &=& 
 \frac{\mbox{w}}{2}   N_2  (1-i\alpha)E_2 -(\bar{\kappa}+i\bar{\omega})E_2  
-(\frac{s}{2}-i\frac{\Omega}{2})E_1
\\  
\frac{d}{dt} N_2 &=& \frac{\mu}{2} - \gamma N_2 - \Gamma (N_2 - N_1) 
- 2\mbox{w} E_2^*E_2 N_2, 
\end{eqnarray}
\end{mathletters}
where $E_1,E_2$ are the field amplitudes of the individual lasers, normalized
so that the intensities $E_1^*E_1, E_2^*E_2$ correspond to photon numbers in 
the cavity and $N_1, N_2$ are the carrier densities above transparency 
in the active media of laser one and two, respectively, each normalized to represent 
the actual number of carriers in the laser. 
The total carrier injection rate $\mu$ is split
equally between the two carrier pools. The linear gain coefficient $\mbox{w}$ is 
then equal to the spontaneous emission rate into the laser mode. The rate
of carrier losses due to electron-hole recombinations by spontaneous emission
into modes not confined to the cavity is given by $\gamma$. The 
properties of the individual laser cavities are given by the
frequency at transparency $\bar{\omega}$ and the cavity loss rate
$\bar{\kappa}$. The carrier density dependent frequency change is 
represented by the linewidth enhancement factor $\alpha$. 

The coupling between the two lasers is described in terms of the parameters 
$\Gamma, s$ and $\Omega$.  The carrier diffusion rate $\Gamma$ represents diffusive
carrier exchange between the two lasers which tends to equilibrate the carrier densities
in the two regions of the gain medium. $\Gamma$ may be derived from the 
ambipolar diffusion constant of the carriers, $D_{diff}$, and the distance
$r$ between the two lasers, resulting in
\begin{equation}
\label{eq:diff}
\Gamma = 4\pi^2 \frac{D_{\mbox{diff}}}{r^2}.
\end{equation}
Note that the consideration of carrier diffusion effects represents an extension 
of the original model of Winful and  Wang \cite{Hof98,Win88}.
The dissipative optical coupling term $s$ represents additional optical
losses in the region between the two lasers, where the two field modes
interfere. Constructive interference ($E_1$ has the same phase as $E_2$) thus leads to
increased losses, while destructive interference (phase difference of $\pi$
and, therefore, opposite sign) leads to reduced losses. This coupling mechanism
is itself sufficient to explain anti-phase locking \cite{Jan97,Ore92}. The
particular physical explanation associated with this term in a real laser device
depends on the type of waveguiding mechanisms used to separate the two lasers. In gain
guided twin stripe arrays, the additional losses are due to internal absorption in the
region of negative gain between the lasers.  In vertical cavity surface emitting laser
arrays fabricated by destroying the Bragg mirrors between the lasers \cite{Ore92}, this
term represents the increased light emission from the cavity in that region.
The coherent coupling of the two localized field modes by optical diffraction is
represented by $\Omega$, the frequency difference between the symmetric and the
anti-symmetric supermodes in the cavity at $N_1 = N_2$.

Since this model represents a perfectly symmetric array of two lasers,
it is useful to express the dynamics in terms of the symmetric and the
anti-symmetric supermode, $E_+ = 1/\sqrt{2}(E_1+E_2)$ and 
$E_- =1/\sqrt{2}(E_1-E_2)$ respectively. Also, the carrier densities can be expressed
in terms of the total carrier density $N=N_1+N_2$ and the carrier density
difference $\Delta = N_1 - N_2$. The equations then read
\begin{mathletters}
\begin{eqnarray}
\label{eq:reqn}  
\frac{d}{dt} E_+ &=& 
 \frac{\mbox{w}}{2}   N  (1-i\alpha)E_+ -(\kappa_++i\omega_+)E_+ 
+\frac{\mbox{w}}{2}\Delta(1-i\alpha)E_- \\  
\frac{d}{dt} E_- &=& 
 \frac{\mbox{w}}{2}   N  (1-i\alpha)E_- -(\kappa_-+i\omega_-)E_- 
+\frac{\mbox{w}}{2}\Delta(1-i\alpha)E_+ \\  
\frac{d}{dt} N &=& \mu - \gamma N - \mbox{w}(E_+^*E_++E_-^*E_-)   N  
                                  - \mbox{w}(E_+^*E_-+E_-^*E_+)\Delta \\
\frac{d}{dt} \Delta &=& -(\gamma+2\Gamma)\Delta - \mbox{w}(E_+^*E_++E_-^*E_-)\Delta
                                                - \mbox{w}(E_+^*E_-+E_-^*E_+)   N,
\end{eqnarray}
\end{mathletters}
where $\kappa_\pm = \bar{\kappa} \pm s/2$ and $\omega_\pm = \bar{\omega}
\pm \Omega/2$. In this formulation, the rate equations of the two laser
array appear as rate equations of a single two mode laser with the additional
variable $\Delta$ describing the spatial holeburning. 

\subsection{Fluctuation dynamics}

If the effects of spatial holeburning are neglected by omitting all
terms which depend on $\Delta$, the stable solution is given by $E_+=0$, since
$\kappa_+ > \kappa_-$. Therefore, only the anti-symmetric supermode
will show laser activity. In the presence of a non-vanishing spatial holeburning
variable $\Delta$, the stability analysis becomes more involved,
as will be discussed below. However, there is always a stationary solution with
\begin{eqnarray}
\Delta&=&0
\nonumber \\ 
N&=&2\kappa_-/\mbox{w}
\nonumber \\
E_+&=&0
\nonumber \\
E_-&=&\sqrt{\frac{\mu}{4\kappa_-}-\frac{\gamma}{2\mbox{w}}} 
\exp [-i(\omega_-+\alpha\kappa_-)t] \nonumber \\
    &=&\sqrt{I_0} \exp [-i(\omega_-+\alpha\kappa_-)t],
\end{eqnarray}
where $I_0$ is the total laser intensity in terms of the photon number
in the anti-symmetric supermode. 
The linearized dynamics of small fluctuations around this stationary 
solution may be separated into the relaxation oscillations of the 
laser mode and the fluctuations in the non-lasing mode and the holeburning
parameter. This independence of the linear dynamics in $E_+$ and $\Delta$ 
from fluctuations
in $E_-$ and $N$ is a consequence of the fact that $E_-=\Delta=0$ is stationary
for all values of $E_+$ and $N$. Only the relaxation and oscillation rates of
the small fluctuations in $E_+$ and $\Delta$ are altered by $E_+$ and $N$. 
Therefore, the fluctuations in $E_+$ and $N$ appear only in the quadratic terms
of the fluctuation dynamics, as products with $\Delta$ and $E_+$, 
respectively.

While the linearized fluctuations in the lasing mode describe the unaltered
relaxation oscillations of a single mode laser, the fluctuations in the
non-lasing mode are described by
\begin{mathletters}
\begin{eqnarray}  
\frac{d}{dt} E_+ &=& 
 -(\kappa_+-\kappa_-+i\omega_++i\alpha\kappa_-)E_+ 
+\frac{\mbox{w}}{2}\Delta(1-i\alpha)\sqrt{I_0}\; e^{-i(\omega_-+\alpha\kappa_-)t}\\  
\frac{d}{dt} \Delta &=& -(\gamma+2\Gamma+\mbox{w}I_0)\Delta 
          - 2\kappa_-\sqrt{I_0}(e^{-i(\omega_-+\alpha\kappa_-)t}E_+^*
                               +e^{+i(\omega_-+\alpha\kappa_-)t}E_+).
\end{eqnarray}
\end{mathletters}
Note that the fluctuation dynamics can
be written using $\delta E_+ = E_+$ and $\delta \Delta = \Delta$
because the averages of the non-lasing field mode $E_+$ and the 
holeburning parameter$\Delta$ are zero. 

Both the amplitude  $\sqrt{I_0}$ and the phase
$-(\omega_-+\alpha\kappa_-)t$ of the average laser field appear
in the fluctuation dynamics, representing the hole burning effects of
the interference term between the weak non-lasing mode and the laser 
field. Thus the fluctuation dynamics are not independent of
the phase of $E_+$. The phase sensitivity can be represented in the
fluctuation dynamics
by expressing the complex field amplitude $E_+$ in terms of 
the component $f_\parallel$ in phase with the laser field and 
the component $f_\perp$ which is $\pi/2$ out of phase with the laser field,
i.e.
\begin{equation}
E_+ = (f_\parallel-i f_\perp) \exp [-i(\omega_-+\alpha\kappa_-)t].
\end{equation}

The fluctuation dynamics can now be rewritten as a linear matrix equation
in a real three dimensional space. Adding the quantum noise terms $Q_\perp$
and $Q_\parallel$, one obtains the linearized Langevin equation of the 
fluctuations in the non-lasing mode,
\begin{equation}
\label{eq:Lang}
\frac{d}{dt}
\left( \begin{array}{c}
f_\parallel\\f_\perp\\ \Delta 
\end{array} \right)
=
\left(\begin{array}{ccc}
        -s           & +\Omega &      \mbox{w}\sqrt{I_0}/2   \\
     -\Omega         &    -s   & \alpha \mbox{w}\sqrt{I_0}/2 \\
-4\kappa_-\sqrt{I_0} &     0   & -\gamma-2\Gamma-\mbox{w}I_0      
\end{array}\right)               
\left(\begin{array}{c}f_\parallel\\f_\perp\\ \Delta\end{array}\right)
+
\left(\begin{array}{c}Q_\parallel\\Q_\perp\\ 0\end{array}\right),
\end{equation}
The quantum noise terms include quantum fluctuations of the light field
vacuum entering the cavity and the dipole fluctuations of the gain 
medium. Here we have neglected the shot noise terms acting on the carrier 
density difference
$\Delta$. This assumption can be well justified in the case of 
$\bar{\kappa} > \gamma$, where within an order of
magnitude of the threshold current the total number of carriers in the gain
medium is much larger than the number of photons in the cavity . The relative 
shot noise in the carrier system will
therefore be much lower than the photonic shot noise in the light field system. 
Note that the presence of relaxation oscillations is an indicator that this 
assumption is valid. The limit of its validity is reached when the
stimulated emission overdamps the relaxation oscillations.

The quantitative expression for the optical quantum noise terms $Q_\parallel$
and $Q_\perp$ may be derived by comparing the corresponding intensity
noise with the expected shot noise for the quantized processes. The vacuum
noise entering the cavity thus corresponds to the quantization of photons
emitted from the cavity and the dipole noise in the gain medium corresponds 
to the quantization of
photon emission from the gain medium into the cavity. Note that 
reabsorption of photons due to incomplete inversion may add extra noise.
In the following, however, we assume perfect inversion in the gain medium 
by using the minimum noise terms required by quantum mechanics,
\begin{equation}
\langle Q_\parallel(t)Q_\parallel(t+\tau) \rangle =
\langle Q_\perp(t)Q_\perp(t+\tau) \rangle = 
\kappa_-\delta(\tau).
\end{equation}

The fluctuation dynamics described by equation (\ref{eq:Lang}) depends on
two relaxation times and two oscillation frequencies. In the case of weak
laser light ($I_0\approx 0$), the light field dynamics are governed by the
frequency $\Omega$ representing the frequency difference between the 
supermodes and the damping coefficient $s$ representing the loss rate 
difference between the supermodes. The interference effects between the
laser field and the light field fluctuations in the non-lasing mode 
couple the holeburning parameter $\Delta$ to the field dynamics of the
non-lasing mode. The total coupling strength is given by the relaxation
oscillation frequency $\sqrt{2\kappa_-\mbox{w}I_0}$. The damping of the holeburning
parameter $\Delta$ is given by the sum of the spontaneous emission rate
$\gamma$, the Diffusion rate $2\Gamma$ and the rate of stimulated emission
$\mbox{w}I_0$. 

To illustrate the physical processes involved in the coupling between the spatial
holeburning parameter $\Delta$ and the field fluctuations in the 
non-lasing mode, it is useful to examine the real time dynamics of the
fluctuations in $f_\parallel$ and $f_\perp$ for large intensities $I_0$.
If $\sqrt{2\kappa_-\mbox{w}I_0}\gg \Omega$, relaxation oscillations dominate the 
dynamics of the fluctuations. However, since the holeburning is only caused 
by the field fluctuation $f_\parallel$ which are in phase with the laser field,
any fluctuations in $f_\perp$ will not cause oscillations. Instead, 
$f_\perp$ relaxes exponentially with
\begin{eqnarray}
\label{eq:frel}
\Delta &\approx& f_\parallel \approx 0 \nonumber \\
f_\perp &\approx& f_0 e^{-(s+\alpha\Omega)t}.
\end{eqnarray}
The effect of the frequency difference 
$\Omega$ is suppressed by the fast relaxation oscillations and would not have
any influence on the dynamics if it were
not for the phenomena expressed by the linewidth enhancement factor
$\alpha$. The $\alpha$ factor describes waveguiding properties induced
by spatial holeburning. If a carrier density difference
$\Delta$ exists between the two cavities, $\alpha$ induces a change in the relative
phase between the local fields $E_1$ and $E_2$. In terms of the supermodes,
this means that the components of $E_-$ and $E_+$ which are out of phase
by $\pi/2$ are coupled. Therefore, $\Delta$ causes a change in $f_\perp$
via $\alpha$ by coherently converting part of the laser light amplitude
into the non-lasing mode. In the presence of a strong laser field, even 
the negligibly small holeburning caused as $\Omega$ rotates $f_\perp$ into 
$f_\parallel$ is sufficient to diminish $f_\perp$ by destructive interference 
with the light coherently transferred from the laser mode.

In the relaxation oscillations caused by the spatial holeburning effect of
interference between the laser light and in-phase fluctuations in the 
non-lasing modes, the $\alpha$ factor produces an out-of-phase component
of $\alpha$ times the in-phase fluctuation, i.e. $f_\perp = \alpha f_\parallel$.
Also, some of the field in $f_\perp$ is rotated into $f_\parallel$ by the
frequency difference $\Omega$ to cause an effective undamping of the
oscillations. Note that this is the exact counterpart of the additional 
damping in the exponential decay of $f_\perp$. The real time evolution
of the relaxation oscillations reads
\begin{eqnarray}
\label{eq:fdyn}
f_\parallel&\approx&f_0 \cos (\sqrt{2\kappa_-\mbox{w}I_0}t) 
                e^{-(\gamma+2\Gamma+\mbox{w}I_0+s-\alpha\Omega)t/2} \nonumber \\
f_\perp&\approx&\alpha f_0 \cos (\sqrt{2\kappa_-\mbox{w}I_0}t) 
                e^{-(\gamma+2\Gamma+\mbox{w}I_0+s-\alpha\Omega)t/2} \nonumber \\
\Delta &\approx& -2\sqrt{\frac{2\kappa_-}{\mbox{w}}}f_0 \sin (\sqrt{2\kappa_-\mbox{w}I_0}t) 
                e^{-(\gamma+2\Gamma+\mbox{w}I_0+s-\alpha\Omega)t/2}.
\end{eqnarray}    
If the dynamics are dominated by the relaxation oscillations, the exponentially damped
dynamics of $f_\perp$ should produce a Lorentzian line at the laser frequency with a 
width of $2s+2\alpha\Omega$ in the spectrum of the non-lasing mode,
while the relaxation oscillation  dynamics should form two sidebands shifted from the
laser line by the relaxation oscillation frequency and having a width of
$\gamma+2\Gamma+\mbox{w}I_0+s-\alpha\Omega$. If the $\alpha$ factor was zero, these
sidebands would be fully symmetric. The main effect of $\alpha$ is to correlate
the fluctuations in $f_\parallel$ with the fluctuations in $f_\perp$ as expressed in
equation (\ref{eq:fdyn}). If the relaxation starts at $f_\parallel = f_0$ and
$f_\perp = 0$, the total relaxation dynamics are given by
\begin{eqnarray}
\label{eq:asym}
f_\parallel (t) &=&
\cos (\sqrt{2\kappa_-\mbox{w}I_0}t) e^{-(\gamma+2\Gamma+\mbox{w}I_0+s-\alpha\Omega)t/2} \nonumber \\
f_\perp (t) &=&
\alpha \left(\cos (\sqrt{2\kappa_-\mbox{w}I_0}t) e^{-(\gamma+2\Gamma+\mbox{w}I_0+s-\alpha\Omega)t/2}
- e^{-(s+\alpha\Omega)}) \right).
\end{eqnarray}
Figure \ref{Fig1} shows this relaxation trajectory for a typical choice of laser
parameters. The question of whether this trajectory contributes to a high frequency line or
to a low frequency line depends on its curvature. The parts of the trajectory curved left
rotate counterclockwise  and contribute mainly to the low frequency line, while those parts
curved right rotate clockwise and therefore contribute mainly to the high frequency line.
Note that the sections of the trajectory contributing mainly to the low frequency sideband
are significantly longer than the sections contributing mainly to the high frequency 
sideband. It is therefore to be expected that the high frequency sideband is lower
than the low frequency sideband because of the $\alpha$ factor correlation
between the relaxation oscillations along $f_\perp=\alpha f_\parallel$ and the exponential
relaxation along $f_\parallel=0$. 

If $\alpha\Omega$
is larger than $\gamma+2\Gamma +s$ the relaxation oscillations may become undamped
for intensities $\mbox{w}I_0 < \alpha\Omega-\gamma-2\Gamma - s$. However, as the relaxation
oscillations are slower at low intensities $\mbox{w}I_0$, this effect does not
necessarily occur at
$\mbox{w}I_0 = \alpha\Omega-\gamma-2\Gamma - s$ as predicted by equation (\ref{eq:fdyn}).
To find the correct stability boundaries, we will in the following examine the stability
condition of the complete fluctuation dynamics given by ${\bf S}$. 

\subsection{Stability Analysis}

The stability boundaries associated with equation (\ref{eq:Lang}) can be
determined from the matrix 
\begin{equation}
{\bf S} =
\left(\begin{array}{ccc}
         s           & -\Omega &       - \mbox{w}\sqrt{I_0}/2 \\
      \Omega         &     s   & -\alpha \mbox{w}\sqrt{I_0}/2 \\
 4\kappa_-\sqrt{I_0} &     0   &   \gamma+2\Gamma+\mbox{w}I_0      
\end{array}\right)  
\end{equation}
describing the linearized dynamics.
At the stability boundaries, ${\bf S}$ must have at least one eigenvalue
with a real part of zero. This condition is satisfied if there exists a 
frequency  $\omega$ with
\begin{equation}
\label{eq:det}
det\{{\bf S}-i\omega\} = 0.
\end{equation}
The stability boundary for non-zero $\omega$ corresponding to the undamping
of relaxation oscillations by $\alpha\Omega$ can be found by separating the
real and imaginary part of equation (\ref{eq:det}) and eliminating $\omega$.
The condition for stability is then found to be
\begin{equation}
\label{eq:stability}
\alpha\Omega < s+\gamma+2\Gamma+\mbox{w}I_0 + \frac{s}{\kappa_- \mbox{w}I_0}
\left(\Omega^2 +
(s+\gamma+2\Gamma +\mbox{w}I_0)^2\right).
\end{equation}
For the case of $s=0$ equation (\ref{eq:stability}) reduces to the expressions 
derived in previous studies \cite{Hof98,Win88}.
Since an instability is most likely to occur at low intensities $I_0$,
however, a closer examination of the stability boundary is of interest
in the present context. 
In particular, it is possible to determine the 
condition for no instability at any laser intensity,
\begin{equation}
\label{eq:stabcon}
\gamma + 2\Gamma + s > \alpha\Omega \left(1+2\frac{s}{\kappa_-}
- 2 \sqrt{\frac{s}{\kappa_-}
(1+\frac{s}{\kappa_-})(1+\frac{1}{\alpha^2})}\right).
\end{equation}
If the device parameters satisfy this condition, the anti-phase locked operation
of the two laser array is stable over the whole operating range described by
the two density model.

\section{Spectrum of the non-lasing supermode}
\label{sec:spectrum}
\subsection{General formula}
The Langevin equation (\ref{eq:Lang}) can be solved analytically in the frequency 
regime by determining the Greensfunction corresponding to the dynamical matrix ${\bf S}$
and applying it to the white noise input. The Greensfunction is obtained by inverting
the matrix ${\bf S}+i\omega$. In the two dimensional subspace describing only the 
optical field components $f_\parallel$ and $f_\perp$, the Greensfunction of the two 
density model is
\[
{\bf G}(\omega) = \frac{1}{(s+i\omega)^2+\Omega^2+M(\alpha\Omega+s+i\omega)}
\left( \begin{array}{cc}
s+i\omega & \Omega \\ -\Omega-\alpha M & s+i\omega + M
\end{array}\right)
\]
\begin{equation}
\label{eq:green}
\mbox{with}\hspace{1cm} M = \frac{2\kappa_- \mbox{w}I_0}{\gamma+2 \Gamma +\mbox{w}I_0 + i\omega}
\end{equation}
Note that $M$ is a complex function of $\omega$ and $I_0$ which includes all
parameters associated with the carrier dynamics. For $M=0$, the
Greensfunction is that of a cavity mode with a frequency of $-\Omega$ relative to the
laser  line and a damping rate of $s$. $M$ increases with increasing laser intensity $I_0$,
describing the transition from the cavity resonance to relaxation oscillations.

Since the noise input given by $Q_\parallel$ and $Q_\perp$ is white noise of intensity
$\kappa_-$ in both components, the resulting fluctuations are
\begin{equation}
\left(
\begin{array}{cc}
\langle \mid f_\parallel (\omega) \mid ^2 \rangle & 
\langle f_\perp^*(\omega) f_\parallel (\omega) \rangle
\\
\langle f_\parallel (\omega)^* f_\perp (\omega) \rangle &
\langle \mid f_\perp (\omega) \mid ^2 \rangle
\end{array}
\right)
= \frac{\kappa_-}{2\pi} {\bf G}(\omega) {\bf G}^\dagger (\omega).
\end{equation}
The spectrum is given by the average intensities of the field as a function of the
frequency $\omega$. It can therefore be determined from the fluctuations in $f_\parallel$
and $f_\perp$ using
\begin{eqnarray}
I_+(\omega) &=& \langle E_+^*(\omega)E_+(\omega) \rangle \nonumber \\
          &=& \langle \mid f_\parallel (\omega) \mid ^2 \rangle
             +\langle \mid f_\perp (\omega) \mid ^2 \rangle
             + i \langle f_\parallel^*(\omega) f_\perp (\omega) \rangle
             - i \langle f_\perp^*(\omega) f_\parallel (\omega) \rangle.
\end{eqnarray}
The spectrum of the non-lasing symmetric supermode $I_+(\omega)$ for any given set
of cavity parameters $\Omega, s$ and any complex holeburning function $M$ is then
given by the general formula
%
\small
\begin{eqnarray}
\label{eq:spec}
& & I_+(\omega) = \frac{\kappa_-}{2\pi}\\ 
&\times&
\left(
    \frac{2(s^2+(\omega-\Omega)^2)
          +(1-i\alpha)(s-i(\omega-\Omega))M
          +(1+i\alpha)(s+i(\omega-\Omega))M^*
          +(1+\alpha^2)M^*M}
         {\mid (s+i\omega)^2 + \Omega^2 + M (s+i\omega+\alpha\Omega) \mid ^2}\right).\nonumber
\end{eqnarray}
\normalsize

\subsection{General properties of the spectrum}

To understand the physical processes associated with the spectrum described by 
the general formula given in equation
(\ref{eq:spec}), it is helpful to discuss the two limiting cases given by $M=0$ and
$M\rightarrow\infty$. For $M=0$, equation (\ref{eq:spec}) describes the amplified
spontaneous emission spectrum expected without spatial holeburning at a total carrier
density pinned to $N=2\kappa_-/\mbox{w}$,
\begin{equation}
I_+(\omega)_{M=0}= \frac{\kappa_-}{\pi}\left(\frac{1}{s^2+(\omega+\Omega)^2}\right).
\end{equation}
Here, the total photon number in the non-lasing cavity mode is given by $\kappa_-/s$.
In this case, no phase relation exists between the non-lasing mode and the lasing mode.
In the opposite limit, i.e. $M \rightarrow \infty$, only the component of amplified
spontaneous emission  out of phase with the laser mode remains undamped by the high
stimulated emission rate and the spectrum shows a single line at the laser frequency
$\omega=0$,
\begin{equation}
\label{eq:infty}
I_+(\omega)_{M\rightarrow \infty}=
\frac{\kappa_-}{2\pi}\left(\frac{1+\alpha^2}{(s+\alpha\Omega)^2+\omega^2}\right).
\end{equation}
For $M \rightarrow \infty$ the total photon number in the non-lasing supermode
corresponding to this spectrum is
$\kappa_-(1+\alpha^2)/2(s+\alpha\Omega)$. For $\alpha=0$, this intensity is one
half of the intensity at $M=0$, reflecting the suppression of out-of-phase emissions.
The linewidth enhancement factor $\alpha$ changes this simple situation however. It
enhances the noise effect by a factor of $1+\alpha^2$. This is the same mechanism which 
enhances the phase  noise in the laser light and therefore increases the linewidth of
the laser line. Moreover,
$\alpha$ introduces an additional damping effect of $\alpha\Omega$, which makes this 
line wider than the one for $M=0$ and reduces the total photon number accordingly.

The typical spectra pertaining to values of the complex rate $M$ between these two
extremes are triplet spectra. The case described by
equations (\ref{eq:fdyn}) features not only the centerline corresponding to
$M\rightarrow\infty$, but also a pair of relaxation oscillation sidebands corresponding to
the in-phase component of amplified spontaneous  emission. In the limit of fast relaxation
oscillations assumed in equations  (\ref{eq:fdyn}), the sidebands are centered around the
relaxation oscillation frequency
$\omega_R =\pm\sqrt{2\kappa_-\mbox{w}I_0}$. Therefore, the sidebands can be derived from 
equation (\ref{eq:spec}) by
approximating the spectrum for large $\pm\omega_R$ in the vicinity of $\omega=\omega_R$.
The  result of this approximation, including the lowest order asymmetry between the
sidebands, reads
\begin{eqnarray}
\label{eq:trip}
I_+(\omega) &\approx&
\frac{\kappa_-}{2\pi}\left(\frac{1+\alpha^2}{(s+\alpha\Omega)^2+\omega^2}\right)
\nonumber \\ &&
+\frac{\kappa_-}{2\pi}\left(
 \frac{1+\alpha^2 + 2(\Omega+\alpha(s+\gamma+2\Gamma +\mbox{w}I_0))/\sqrt{2\kappa_-\mbox{w}I_0}}
      {(\gamma+2\Gamma+\mbox{w}I_0+s-\alpha\Omega)^2+4(\omega+\sqrt{2\kappa_-\mbox{w}I_0})^2}\right)
\nonumber \\ &&
+\frac{\kappa_-}{2\pi}\left(
 \frac{1+\alpha^2 - 2(\Omega+\alpha(s+\gamma+2\Gamma +\mbox{w}I_0))/\sqrt{2\kappa_-\mbox{w}I_0}}
      {(\gamma+2\Gamma+\mbox{w}I_0+s-\alpha\Omega)^2+4(\omega-\sqrt{2\kappa_-\mbox{w}I_0})^2}\right).
\end{eqnarray} 
This triplet spectrum can be characterized by the total photon numbers in the three
lines,
\begin{eqnarray}
I_{centerline} &=& \frac{(1+\alpha^2)\kappa_-}{2(s+\alpha\Omega)} \nonumber \\
I_{sidebands} &=& \frac{(1+\alpha^2)\kappa_-}{4(\gamma+2\Gamma+\mbox{w}I_0+s-\alpha\Omega)}
\left(1\pm\frac{2\Omega+2\alpha(s+\gamma+2\Gamma+\mbox{w}I_0)}{(1+\alpha^2)\sqrt{2\kappa_-
\mbox{w}I_0}}\right).
\end{eqnarray}
The sidebands are clearly damped by the stimulated emission into the laser line,
$\mbox{w}I_0$, as well as by spontaneous recombinations and carrier diffusion expressed by
$\gamma+2\Gamma$. The increase in the stimulated emission rate is responsible for the
suppression of the sidebands with increasing laser intensity. Note that the
undamping of relaxation oscillations discussed in section \ref{sec:SDM} also appears in
the total intensity of the sidebands, possibly raising the sideband intensities far above
the intensity of the centerline at low laser intensities. 

The asymmetry of the sidebands can be traced back to two separate contributions, one
proportional to the frequency difference between the modes $\Omega$ and
one proportional to $\alpha(s+\gamma+2\Gamma+\mbox{w}I_0)$. In both cases, the low frequency
sideband intensity is greater than the high frequency sideband intensity. The first term
can be understood as a remnant of the  cavity resonance of the non-lasing supermode in the
absence of spatial holeburning. The low frequency line continuously evolves from the 
line at $\omega=-\Omega$ for $M=0$ while the high frequency line only emerges 
as the laser intensity is increased. However, the low frequency line is also stronger
for $\Omega=0$. This is the effect explained in section \ref{sec:SDM}.B which is 
caused by the $\alpha$ factor. Within the approximation assumed here, the sideband
intensity ratio corresponding to the parameters given in Figure \ref{Fig1} is 
about two to one.

While the general considerations given above already cover the qualitative effects
which can be expected in the amplified spontaneous emission spectrum, the quantitative
changes in the spectrum as spatial holeburning effects get stronger with increasing laser
intensity are best illustrated and discussed on spectra pertaining to typical
examples of device parameters. In the following, we will therefore discuss four cases,
with either strong  centerlines ($\Omega<s$) or weak centerline ($\Omega>s$) and either
low carrier diffusion ($\gamma+2\Gamma \ll \kappa_-$) or high carrier diffusion
($\gamma+2\Gamma =
\kappa_-/2$).

\section{Examples of spectra for typical device parameters}
\label{sec:cases}

\subsection{Diffusion rates and array size}
In the following we will discuss spectra assuming diffusion rates of
$\Gamma \approx 5$ GHz to $\Gamma \approx 100$ GHz. Based on 
a typical ambipolar diffusion constant $D_{\mbox{diff}}$ 
in semiconductor lasers of 
about $10\; \mbox{cm}^2\mbox{s}^{-1}$ the diffusion rate $\Gamma$ is related
to the separation $r$ between the lasers according to equation 
(\ref{eq:diff}) by
\begin{equation}
\left(\frac{r}{1 \mu\mbox{m}}\right)^2 \approx \frac{40 \mbox{GHz}}{\Gamma}.
\end{equation}
Thus for GaAs based semiconductor lasers, the diffusion rates discussed 
correspond to arrays with separations between $r=0.6 \mu\mbox{m}$ and
$r=3 \mu\mbox{m}$. 

In this context the separations between the lasers of an array of 
about $r=3 \mu\mbox{m}$ represent
diffusion effects which are much weaker than the spatial hole burning effects.
This situation is to be expected for most of the semiconductor laser arrays
realized today. Note that the spectra for higher diffusion rates and larger
separations between the lasers are qualitatively similar to the examples
given by the diffusion rates $\gamma+2\Gamma = 10$ GHz and 
$\gamma+2\Gamma = 15$ GHz. Consequently, these results also apply to array
structures with much larger separations between the lasers.
Below a separation between the lasers of $r=3 \mu\mbox{m}$ the effects of
diffusion drastically increase. In the following we investigate two
representative cases of such strong diffusion effects with $\gamma+2\Gamma 
= 50$ GHz and $\gamma+2\Gamma = 200$ GHz. This corresponds to
laser separations of about $r=1.25 \mu\mbox{m}$ and $r=0.65 \mu\mbox{m}$,
respectively. These values were chosen in view of a future realization of
highly integrated and/or short wavelength semiconductor laser arrays. 
The spectra for
$\gamma+2\Gamma = 10$ GHz and $\gamma+2\Gamma = 15$ GHz may thus be 
considered typical for the
type of arrays already realized in experiments while the spectra for
$\gamma+2\Gamma = 50$ GHz and $\gamma+2\Gamma = 200$ GHz represent 
possibilities accessible only
if further integration can be achieved.  

\subsection{Triplet structure}
\label{ssec:trip}

A typical triplet structure can be observed in devices where the frequency difference
$\Omega$ is smaller than the dissipative optical coupling $s$. The intensity of the
centerline is then comparable to the intensity of the single line at $M=0$. The
carrier relaxation and diffusion rate $\gamma + 2\Gamma$ will suppress the sidebands, so
a low value for this rate produces the strongest possible sidebands. 
The evolution of the spectrum as a function of laser intensity for the device
parameters $\kappa_-=400$ GHz,
$s=3$ GHz, $\Omega=1$ GHz, $\gamma+2 \Gamma =10$ GHz and $\alpha=3$ is shown in 
Figure \ref{Fig2} and \ref{Fig3}. The laser intensity is given in terms of the stimulated
emission rate $\mbox{w}I_0$ which can be converted to units of threshold current by dividing
$\mbox{w}I_0$ by the spontaneous emission rate $\gamma$. Since $\gamma$ is
usually close to one Gigahertz, realistic values for $\mbox{w} I_0$ range from about zero to ten
Gigahertz.

Figure \ref{Fig2} shows the evolution of the triplet structure with increasing intensity.
Equation (\ref{eq:trip}) describes this evolution very well, except for the immediate
vicinity of the threshold. Note that the asymmetry of the sidebands is extremely strong.
At $\mbox{w}I_0=1$ GHz, the choice of parameters corresponding to figure \ref{Fig1}, the
intensity ratio between the sidebands is about two to one, as predicted by the approximate
equation (\ref{eq:trip}). 

Figure \ref{Fig3} displays an enlarged part of figure \ref{Fig2} close to
threshold. In this region, the triplet has not yet been formed.  Between $\mbox{w}I_0=0$ and
$\mbox{w}I_0=0.03$ GHz the maximum of the single spectral line
rapidly shifts to lower frequencies, increasing
in intensity in the process. Between $\mbox{w}I_0=0.03$ GHz and $\mbox{w}I_0=0.15$ GHz, the line widens
and finally forms a plateau on the low frequency side. At about $\mbox{w}I_0=0.25$ GHz, the
plateau on the low frequency side splits from the center line to form the lower 
sideband. The high frequency sideband emerges in a similar fashion between $\mbox{w}I_0=0.1$
GHz and $\mbox{w}I_0=0.5$ GHz.

\subsection{Sideband suppression}
\label{ssec:center}
Since the center line is only damped by the sum of rates $s+\alpha\Omega$ while
the sidebands are additionally damped by carrier diffusion, the sidebands will be
suppressed in the presence of fast carrier diffusion. In this case, only a single
line which gets frequency locked to the laser frequency as the laser
intensity increases remains in the spectrum. This effect can indeed be seen in the 
spectra by choosing the same parameters as for the triplet  structure (section~\ref{ssec:trip}),
$\kappa_-=400$ GHz,
$s=3$ GHz, $\Omega=1$ GHz and
$\alpha=3$, except for an increased carrier recombination and diffusion rate of
$\gamma + 2\Gamma=200$ GHz. 

Figure \ref{Fig4} clearly shows the
evolution of a single line. However, the peak frequency is not just reduced to 
zero as the amplified spontaneous emission is phase locked to the laser light
by the increasingly strong spatial holeburning effects. Instead, the peak even shifts
to larger negative frequencies just above threshold. Between $\mbox{w}I_0=0.5$ GHz and 
$\mbox{w}I_0=1.0$ GHz the peak frequency is about 2 GHz, twice as large as the frequency 
at $\mbox{w}I_0=0$. This shift to higher frequency differences between the laser light and
the non-lasing mode is similar to the one observed in the case of low carrier
diffusion discussed above. However, the shift is not as rapid in the high diffusion case
and instead extends to much higher laser intensities $\mbox{w}I_0$.
In the vicinity of $\mbox{w}I_0=0$, an analytical expression may
be  derived for the change in peak position by considering
the coupling terms between the holeburning parameter $\Delta$ and the field components
$f_\parallel$ and $f_\perp$ 
in the dynamical matrix ${\bf S}$ as weak perturbations. The peak position $\omega_p$
is given by the imaginary part of the eigenvalues of ${\bf S}$. Near $\mbox{w}I_0=0$ it shifts 
at a rate of
\begin{equation}
\label{eq:shift}
\frac{d(\omega_p)}{d(\mbox{w}I_0)} = \frac{\alpha\kappa_-}{(\gamma+2\Gamma)}.
\end{equation}
Note that this equation generally applies to all parameter sets. However, in the case of
low carrier diffusion it is only valid extremely close to threshold.
In particular, the shift of $d(\omega_p)/d(\mbox{w}I_0)= 120$ for the low diffusion
case  in figure \ref{Fig2} extends only to $\mbox{w}I_0=0.03$ GHz.In the low 
diffusion case displayed in figure \ref{Fig3} the shift  of $d(\omega_p)/d(\mbox{w}I_0)= 6$ 
is the dominant feature of the spectrum up to intensities of $\mbox{w}I_0=0.4$ GHz, 
and even at $\mbox{w}I_0=2$ GHz the peak of the line is still 1 GHz below the laser 
frequency.
These features in the evolution of the spectrum demonstrate how carrier
diffusion suppresses the spatial holeburning effects. We note that a quantitative
expression for the relative importance of spatial holeburning is given by the complex
rate $M$ introduced in equation (\ref{eq:green}). It is the denominator of $M$ 
which includes the carrier diffusion rate $2\Gamma$. Thus increased carrier 
diffusion reduces the absolute value of
$M$, thereby suppressing spatial holeburning effects.

\subsection{Undamped sidebands}
\label{ssec:side}

If $\alpha\Omega$ is larger than $s+\gamma+2\Gamma$, the undamping effect may cause an
instability as discussed in section \ref{sec:SDM}. If the instability is only narrowly
avoided by choosing parameters close to the stability condition (\ref{eq:stabcon}),
the sidebands should grow large and become narrow as the laser intensity passes the region
of instability in the phase diagram. Figure \ref{Fig5} shows the phase diagram for
variable damping of the holeburning parameter, $\gamma + 2\Gamma$, and
$\kappa_- = 100$ GHz, $s=5$ GHz, $\Omega = 10$ GHz, and $\alpha=3$. The horizontal line
at $\gamma+2\Gamma=15$ GHz shows the section of this phase diagram corresponding to the 
choice of parameters in figure \ref{Fig6}. The sidebands clearly have their maximum at a
stimulated emission rate near $\mbox{w}I_0 = 5$ GHz, the point closest to the unstable region
in the phase diagram. The linewidth of the sidebands in this region is close to
1 GHz, as compared to the linewidth given by $2s=10$ GHz. The maximal value of the
total photon number in the non-lasing mode is roughly 240 at $\mbox{w}I_0=5$ GHz. 
For comparison, the total photon number at $\mbox{w}I_0=0$ is 20 and the centerline photon number
for $\mbox{w}I_0\rightarrow\infty$ is 14.3. The asymmetry of the lines is given by an intensity
ratio of about three to one, e.g. at $\mbox{w}I_0=5$ GHz the photon numbers in the sidebands
are approximately 180 in the low frequency sideband and 60 in the high frequency sideband.

\subsection{Sidebands in the presence of strong carrier diffusion effects}
If the carrier diffusion is increased, the undamping of the sidebands is suppressed.
However, the sidebands still remain the dominant feature of the spectrum if the 
line at $\mbox{w}I_0=0$ is well separated from the laser line by $\Omega>s$. This effect
can be illustrated using the same 
parameters as those in the case of undamped sidebands (section~\ref{ssec:side}), 
$\kappa_- = 100$ GHz, $s=5$ GHz, 
$\Omega =10$ GHz and $\alpha=3$, except for an increased carrier damping and diffusion 
rate of $\gamma + 2\Gamma = 50$ GHz. As shown in figure \ref{Fig7}, the low
frequency sideband continuously evolves from the amplified spontaneous emission line at
threshold as intensity increases. Meanwhile, the weak phase locking effects cause a low
intensity mirror image  line to appear at the corresponding position on the high frequency
side. Note that the frequency separation from the laser line increases from $\Omega=10$
GHz at
$\mbox{w}I_0=0$ to almost $40$ GHz at $\mbox{w}I_0=10$ GHz. The relaxation oscillation frequency at
$\mbox{w}I_0=10$ GHz is about 45 GHz. This indicates that the line at
$\mbox{w}I_0=10$ GHz is actually a relaxation oscillation line created by the carrier dynamics,
while the  suppression of the high frequency sideband is not only an effect of $\Omega$
but also of the $\alpha$ factor as described by equation (\ref{eq:asym}). Indeed, the
approximation given by equation (\ref{eq:trip}) predicts an intensity ratio of twenty
to one between the sidebands for the parameters used here.    

\section{Conclusions}
\label{sec:conclusions}

We have demonstrated that 
the spatial holeburning caused by interference between the laser light in the 
antisymmetric supermode of a two laser array and the amplified spontaneous emission in
the non-lasing symmetric supermode can give rise to relaxation oscillations which
phase-lock the amplified spontaneous emission to the laser light. In the limit of 
high laser intensities, this effect splits the spectrum into a center line at the
laser frequency and two sidebands. The center line represents the out-of-phase component
of the amplified spontaneous emission which does not interfere with the laser light.
The two sidebands correspond to the relaxation oscillations between the in-phase
component of amplified spontaneous emission and the depth of spatial holeburning in the
carrier distribution. 

The sidebands may become undamped by the linewidth enhancement factor
$\alpha$, which describes the conversion of laser light from the antisymmetric mode to the
symmetric mode. This effect can be countered by carrier diffusion, which dampens the
sidebands and may actually suppress them. Depending on the device parameters, it is
therefore possible to alternatively find in the spectrum only the sidebands, 
only  a single center line, or the full triplet. The continuous transition from a 
single line at the frequency of the symmetric
supermode to any of the three possibilities can be described analytically using equation
(\ref{eq:spec}). 

In all cases, the amplified spontaneous emission in the non-lasing supermode is strongly
modified by the spatial holeburning effects. The spectra of the non-lasing modes above
laser threshold are therefore quite different from the spectra at or below threshold
described by linear optics.
It should be possible to observe such spectra experimentally by measuring the spectrum
of anti-phase locked laser arrays in the center of the farfield, where contributions
from the anti-symmetric laser mode is close to zero. The type of spectrum observed will
then allow a determination of the coupling strength and the type of coupling in the 
respective device.

\section*{Acknowledgment} 
We sincerely would like to thank Y.~Yamamoto for his hospitality at Stanford University.


%

%

%
\begin{figure}
\caption{Relaxation of $f_\parallel (t=0) =1, f_\perp (t=0) = 0$ for $\kappa_- = 400$ GHz,
$s=3$ GHz, $\Omega=1$ GHz, $\gamma+2\Gamma =10$ GHz, $\alpha =3$, and $\mbox{w}I_0=1$ GHz.
The oscillations effectively correspond to a mainly counterclockwise rotation.}
\label{Fig1}
\end{figure}
\begin{figure} 
\caption{Amplified 
spontaneous emission spectra for
$\kappa_-=400$ GHz, $s = 3$ GHz, $\Omega = 1$ GHz, $\gamma+2\Gamma=10$ GHz and
$\alpha=3$. (a) shows
the contour plot of the spectrum as a function of laser intensity from
$\mbox{w}I_0=0$ to $\mbox{w}I_0=2$ GHz, while
(b) shows the spectra at $\mbox{w}I_0=0$ (no offset), $\mbox{w}I_0=0.5$ GHz (offset of 6/GHz),
$\mbox{w}I_0=1.0$ GHz (offset of 12/GHz), and $\mbox{w}I_0=1.5$ GHz (offset of 18/GHz).}
\label{Fig2}
\end{figure}
\begin{figure} 
\caption{Amplified 
spontaneous emission spectra close to threshold for the same parameters as in 
figure 1. (a) shows the contour plot of the spectrum as a function of laser intensity
from $\mbox{w}I_0=0$ to $\mbox{w}I_0=0.3$ GHz, while
(b) shows the spectra at $\mbox{w}I_0=0$ (no offset), $\mbox{w}I_0=0.1$ GHz (offset of 3/GHz),
$\mbox{w}I_0=0.2$ GHz (offset of 6/GHz), and $\mbox{w}I_0=0.3$ GHz (offset of 9/GHz).}
\label{Fig3}
\end{figure}
\begin{figure} 
\caption{Amplified 
spontaneous emission spectrum for
$\kappa_-=400$ GHz, $s = 3$ GHz, $\Omega = 1$ GHz, $\gamma+2\Gamma=200$ GHz and
$\alpha=3$. (a) shows
the contour plot of the spectrum as a function of laser intensity from
$\mbox{w}I_0=0$ to $\mbox{w}I_0=2$ GHz, while
(b) shows the spectra at $\mbox{w}I_0=0$ (no offset), $\mbox{w}I_0=0.5$ GHz (offset of 6/GHz),
$\mbox{w}I_0=1.0$ GHz (offset of 12/GHz), and $\mbox{w}I_0=1.5$ GHz (offset of 18/GHz).}
\label{Fig4}
\end{figure}
\begin{figure}
\caption{Stability boundary for $\kappa_-=100$ GHz, $s=5$ GHz, $\Omega=10$ GHz, $\alpha=3$
and variable carrier recombination and diffusion rates $\gamma+2 \Gamma$. The diagonal
line is the approximated boundary for sideband undamping given by
$\gamma+2\Gamma+\mbox{w}I_0+s-\alpha\Omega=0$. The horizontal line at $\gamma+2\Gamma=15$ GHz
represents the choice of parameters in section~4~C and in figure~6.}
\label{Fig5}
\end{figure}
\begin{figure}
\caption{Amplified 
spontaneous emission spectrum for
$\kappa_-=100$ GHz, $s = 5$ GHz, $\Omega = 10$ GHz, $\gamma+2\Gamma=15$ GHz and
$\alpha=3$. (a) shows
the contour plot of the spectrum as a function of laser intensity from $\mbox{w}I_0=0$ 
top $\mbox{w}I_0=10$ GHz, while
(b) shows the spectra at $\mbox{w}I_0=3$ GHz (no offset), $\mbox{w}I_0=4$ GHz (offset of 5/GHz),
$\mbox{w}I_0=5$ GHz (offset of 10/GHz), $\mbox{w}I_0=6$ GHz (offset of 15/GHz), and 
$\mbox{w}I_0=7$ GHz (offset of 20/GHz).}
\label{Fig6}
\end{figure}
\begin{figure}
\caption{Amplified spontaneous emission spectrum for
$\kappa_-=100$ GHz, $s = 5$ GHz, $\Omega = 10$ GHz, $\gamma+2\Gamma=50$ GHz and
$\alpha=3$. (a) shows
the contour plot of the spectrum as a function of laser intensity from $\mbox{w}I_0=0$ 
top $\mbox{w}I_0=10$ GHz, while
(b) shows the spectra at $\mbox{w}I_0=0$ (no offset), $\mbox{w}I_0=2$ GHz (offset of 0.2/GHz),
$\mbox{w}I_0=4$ GHz (offset of 0.4/GHz), $\mbox{w}I_0=6$ GHz (offset of 0.6/GHz),
$\mbox{w}I_0=8$ GHz (offset of 0.8/GHz), and $\mbox{w}I_0=10$ GHz (offset of 1.0/GHz).}
\label{Fig7}   
\end{figure}
%




\begin{thebibliography}{12}

\bibitem{Hof97}
H.~F. Hofmann and O. Hess, 
``Quantum Noise and Polarization Fluctuations in Vertical Cavity Surface Emitting Lasers,'' 
Phys. Rev. A {\bf 56,} 868 (1997).

\bibitem{Jan97}
A.~K.~J. van Doorn, M.~P. van Exter, A.~M. van~der Lee, and J.~P. Woerdman,
 ``Coupled-mode description for the polarization state of a vertical-cavity  semiconductor laser,''
Phys. Rev. A {\bf 55,} 1473 (1997).

\bibitem{Lem97}
H. van der Lem and D. Lenstra,
``Saturation-induced frequency shift in the noise spectrum of a birefringent vertical-cavity
surface emitting laser'',
Opt. Lett. {\bf 22}, 1698 (1997).

\bibitem{Hof98}
H.~F. Hofmann and O. Hess,
``The Split Density Model: A Unified Description of
 Polarization and Array Dynamics for Vertical Cavity Surface Emitting Lasers,'' 
Quant. Semiclass. Opt. {\bf 9}, 749 (1997).

\bibitem{Win88}
H.~G. Winful and S.~S. Wang, 
``Stability of phase locking in coupled semiconductor laser arrays,'' 
Appl. Phys. Lett. {\bf 53}, 1894 (1988).

\bibitem{Wan88}
S.~S. Wang and H.~G. Winful, 
``Dynamics of phase-locked semiconductor laser arrays,'' 
Appl. Phys. Lett. {\bf 52}, 1774 (1988).

\bibitem{Ore92}
M. Orenstein, E. Kapon, J.~P. Harbison, L.~T. Florez, and N.~G. Stoffel,
 ``Large two-dimensional arrays of phase-locked vertical cavity surface emitting lasers,'' 
Appl. Phys. Lett. {\bf 60,} 1535 (1992).

\bibitem{Mor92}
R.~A. Morgan, K. Kojima, T. Mullally, G.~D. Guth, M.~W. Focht, R.~E.
Leibenguth, and M. Asom, 
``High-power coherently coupled 8$\times$8 vertical cavity surface emitting laser array,'' 
Appl. Phys. Lett. {\bf 61,} 1160 (1992).

\bibitem{Mor93}
R.~A. Morgan and K. Kojima, 
``Optical characteristics of two-dimensional coherently coupled vertical-cavity surface-emitting laser
arrays,'' 
 Opt. Lett. {\bf 18,} 352 (1993).

\bibitem{Cat96}
J.~M. Catchmark, L.~E. Rogers, R.~A. Morgan, M.~T. Asom, G.~D. Guth, and D.~N. Christodoulides, 
``Optical Characteristics of Multitransverse-Mode Two-Dimensional Vertical-Cavity Top
Surface-Emitting Laser Arrays,''  
{IEEE} J. Quant. Electr. {\bf 32,} 986 (1996).

\bibitem{Hof98a}
H.F. Hofmann and O. Hess,
``Spontaneous-emission spectrum of the nonlasing supermodes in semiconductor laser arrays'',
Opt. Lett. {\bf 23}, 391 (1998).

\bibitem{Mun97}
M. M\"unkel, F. Kaiser, and O. Hess, 
``Stabilization of spatiotemporally chaotic semiconductor laser arrays by means of delayed optical
feedback,'' 
 Phys. Rev. E. {\bf 56} (1997).

\end{thebibliography}
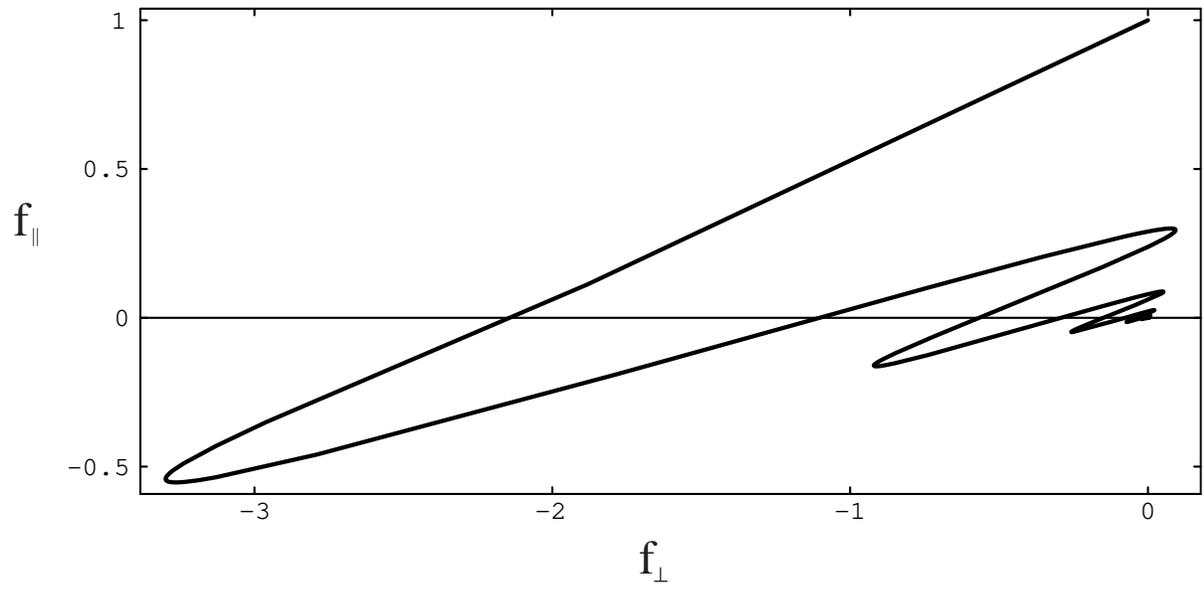
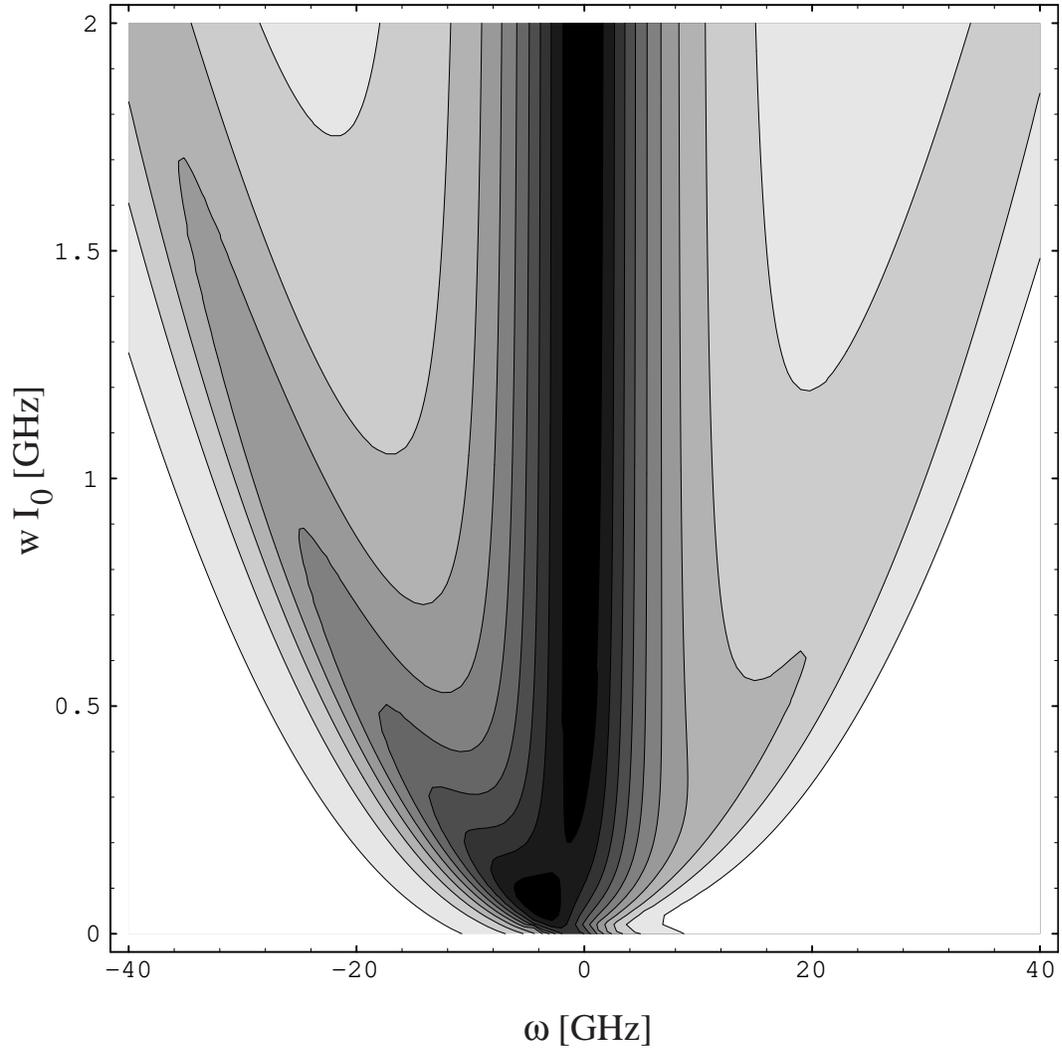
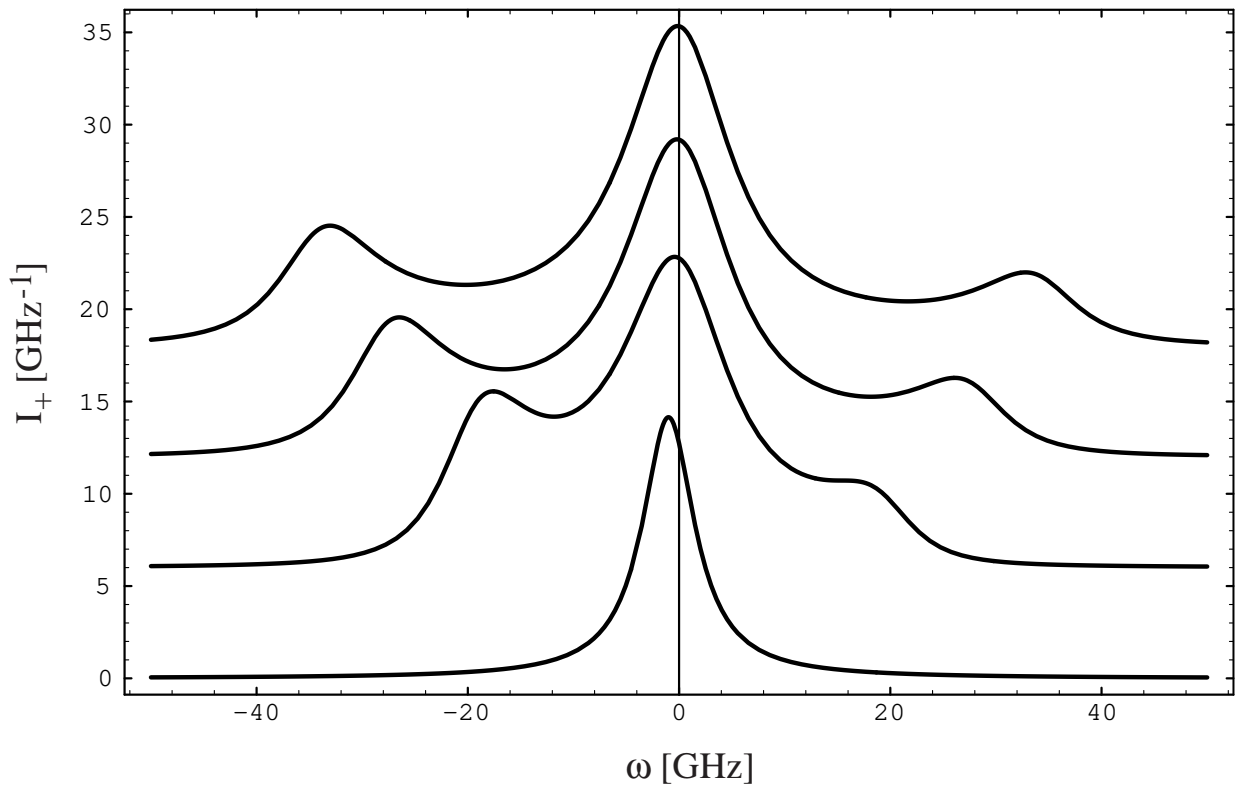
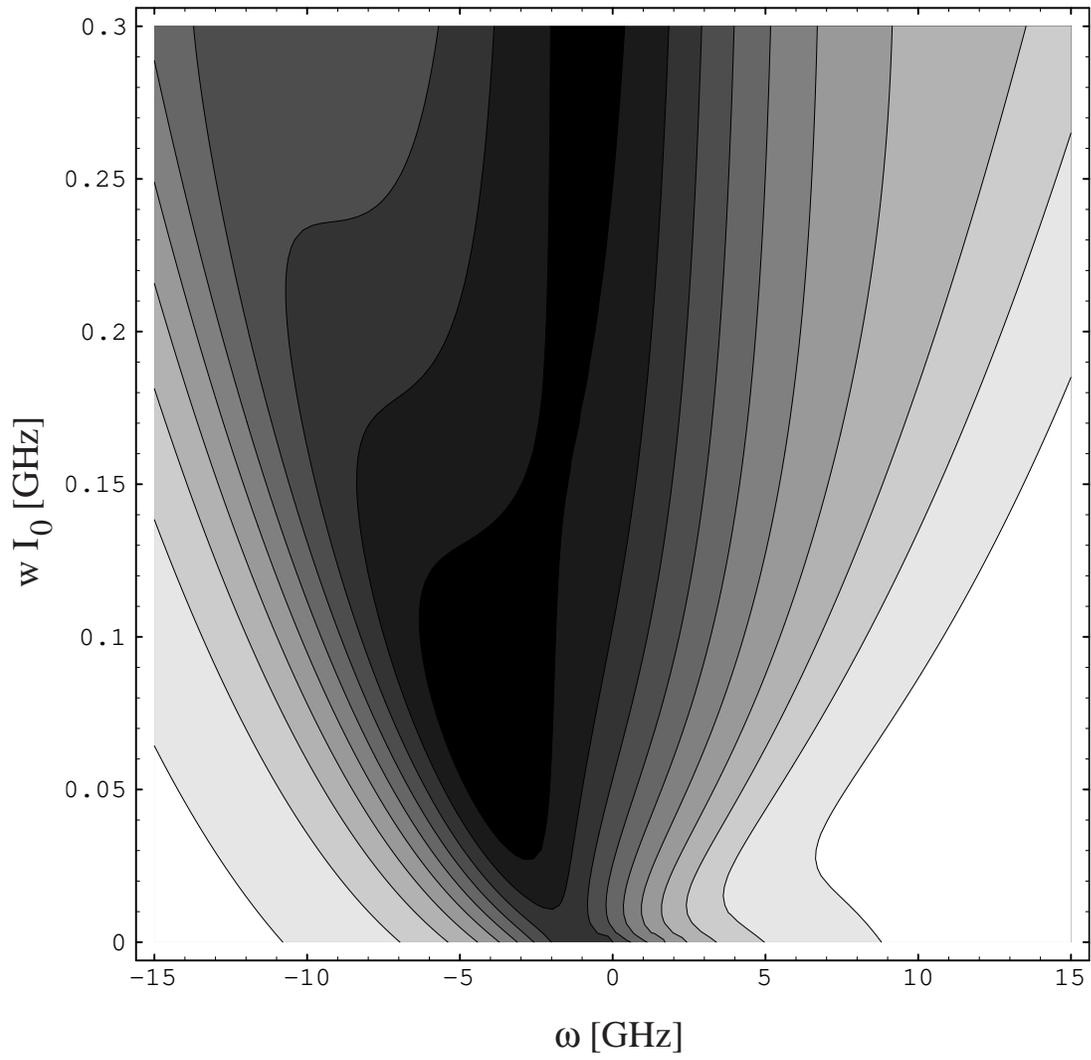
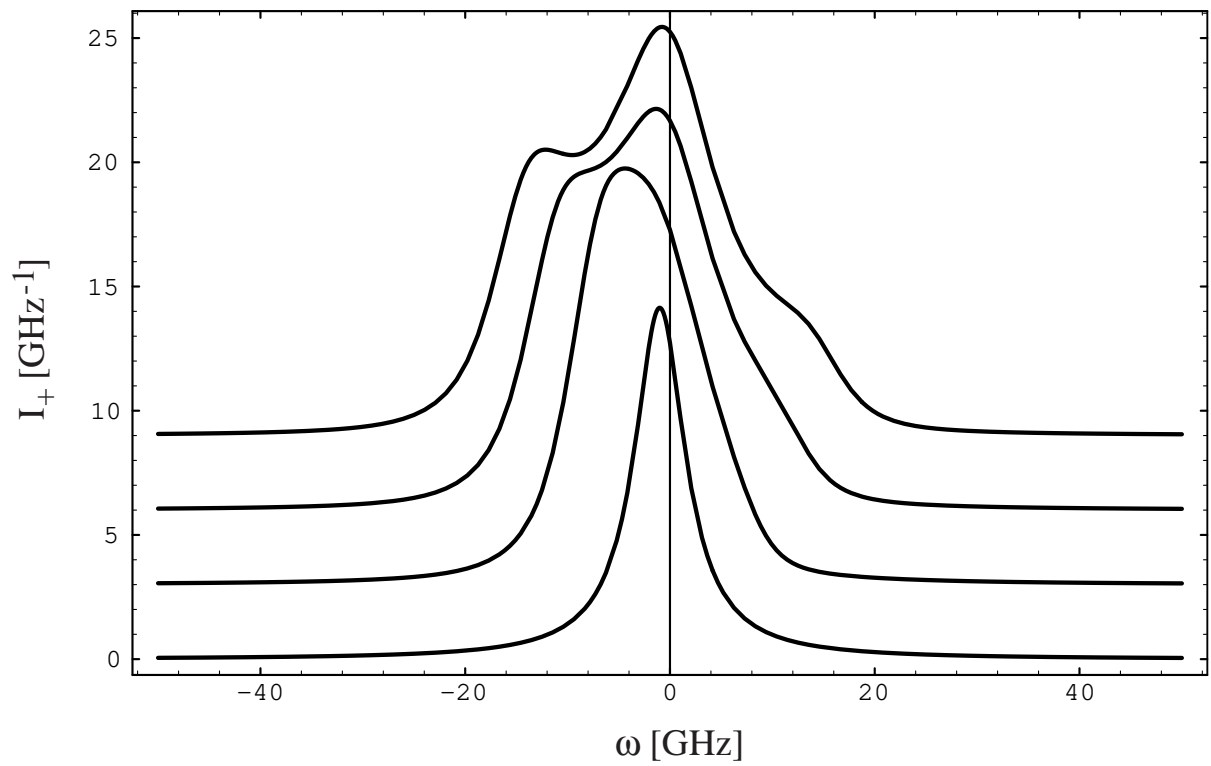
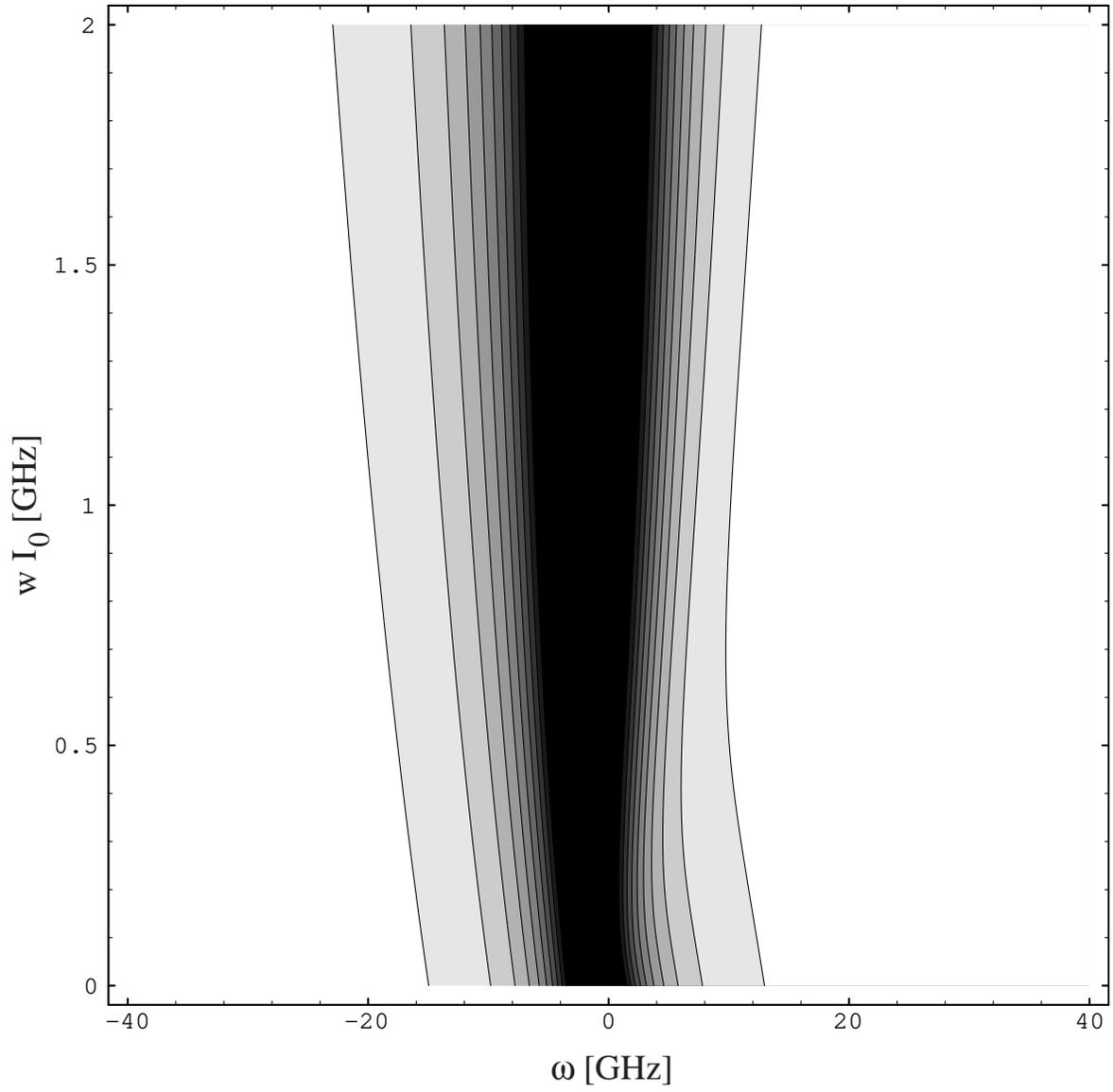
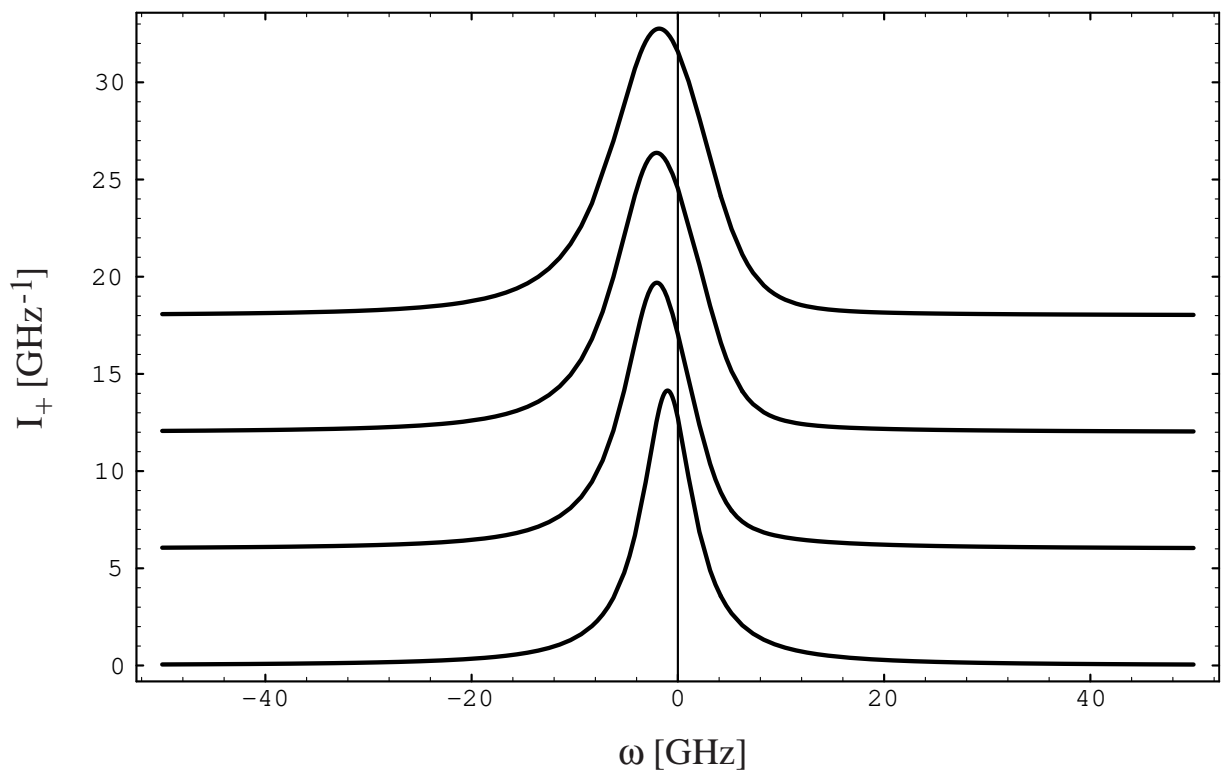
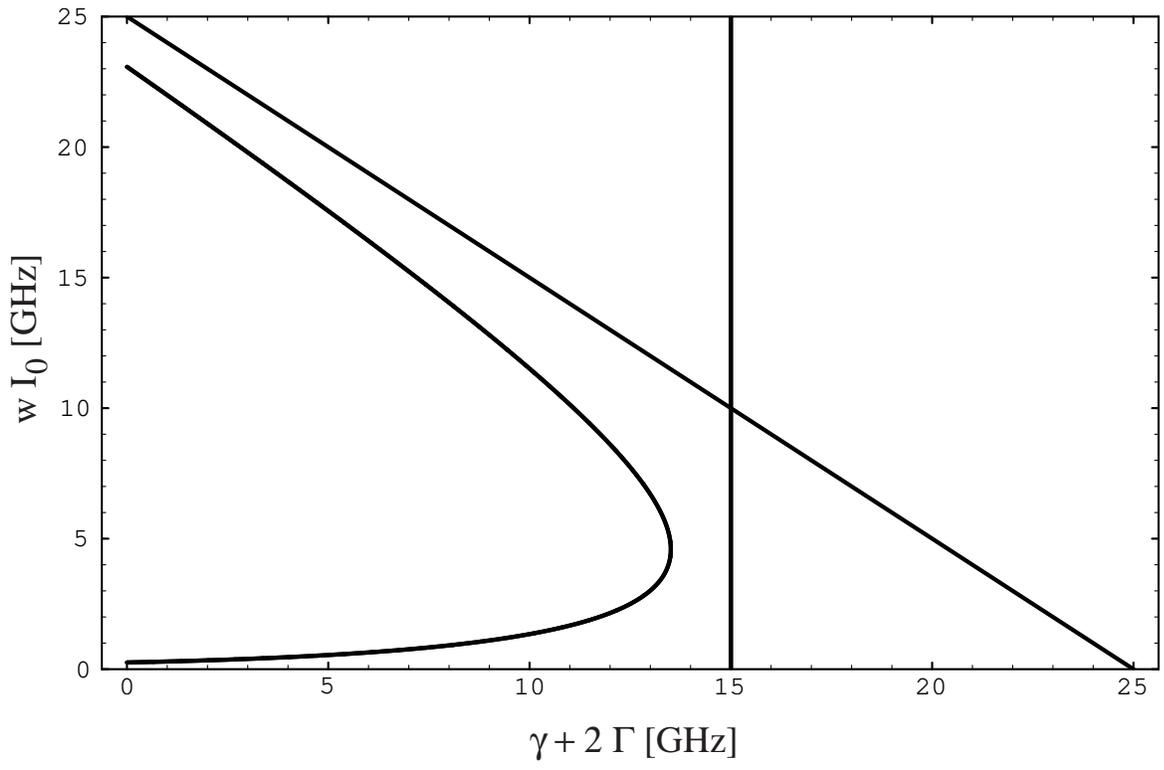
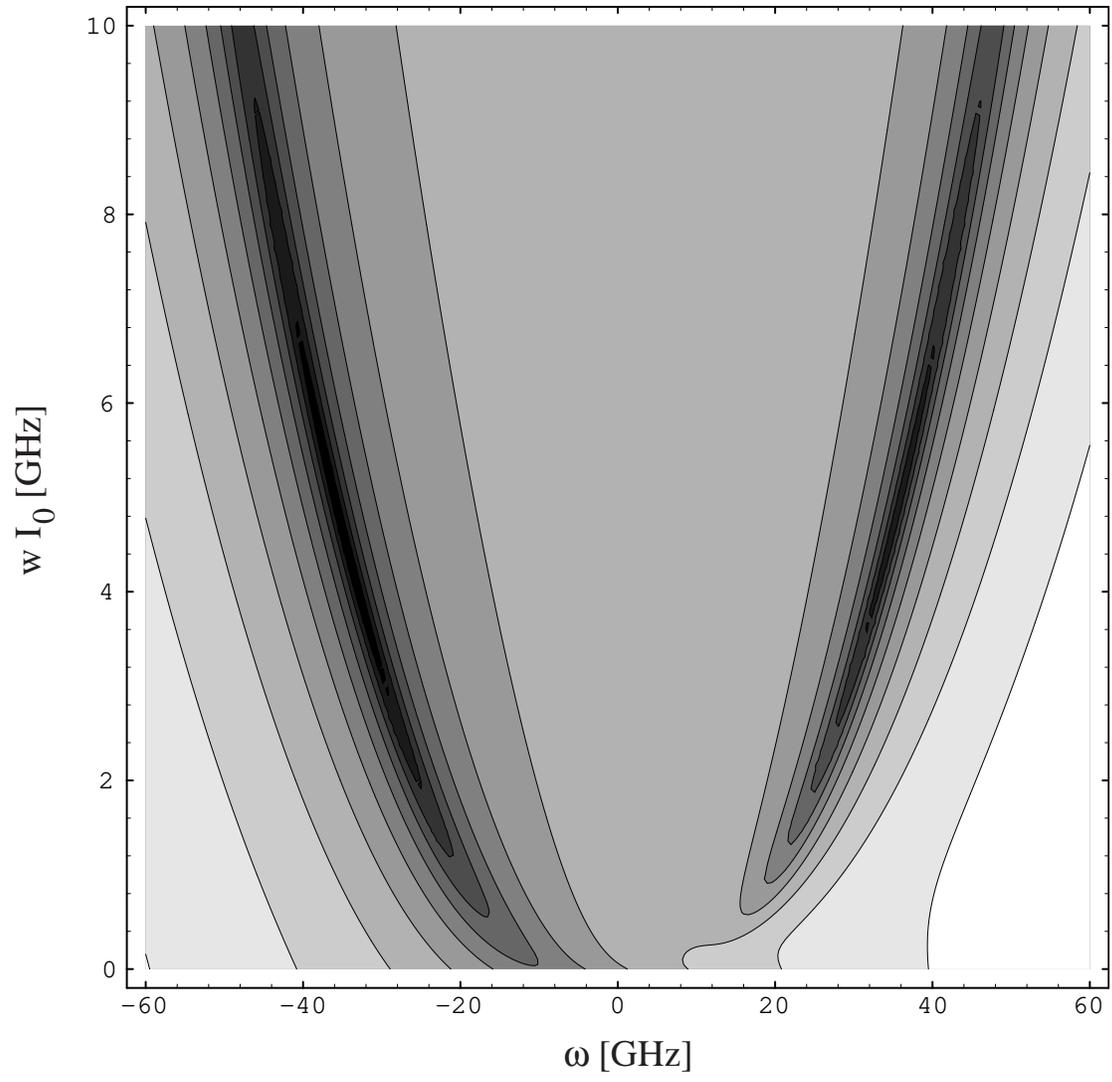
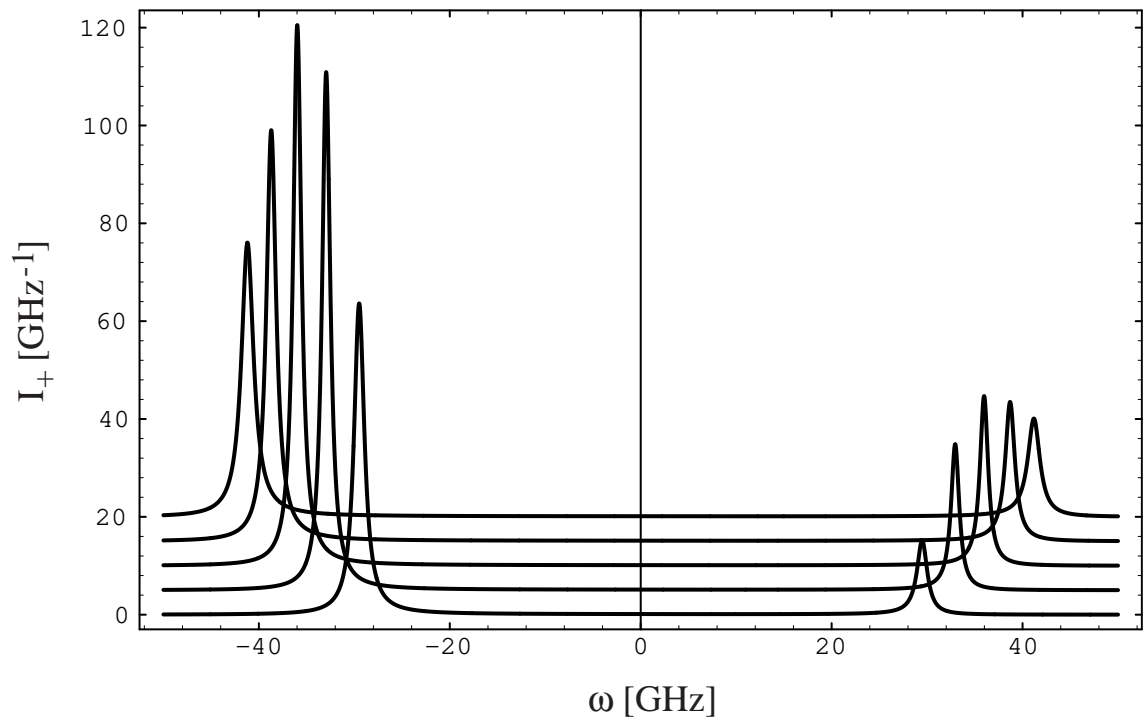
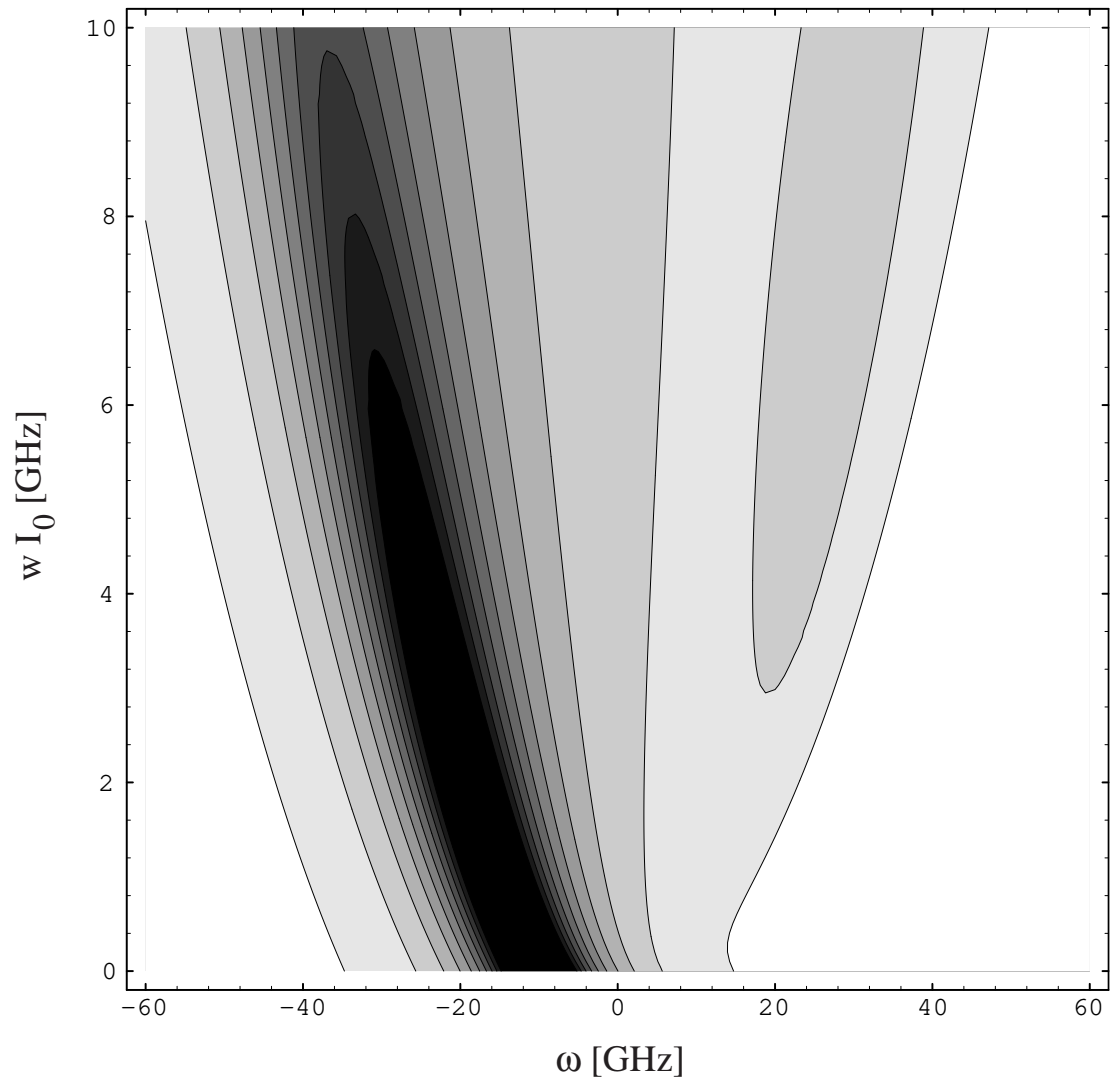
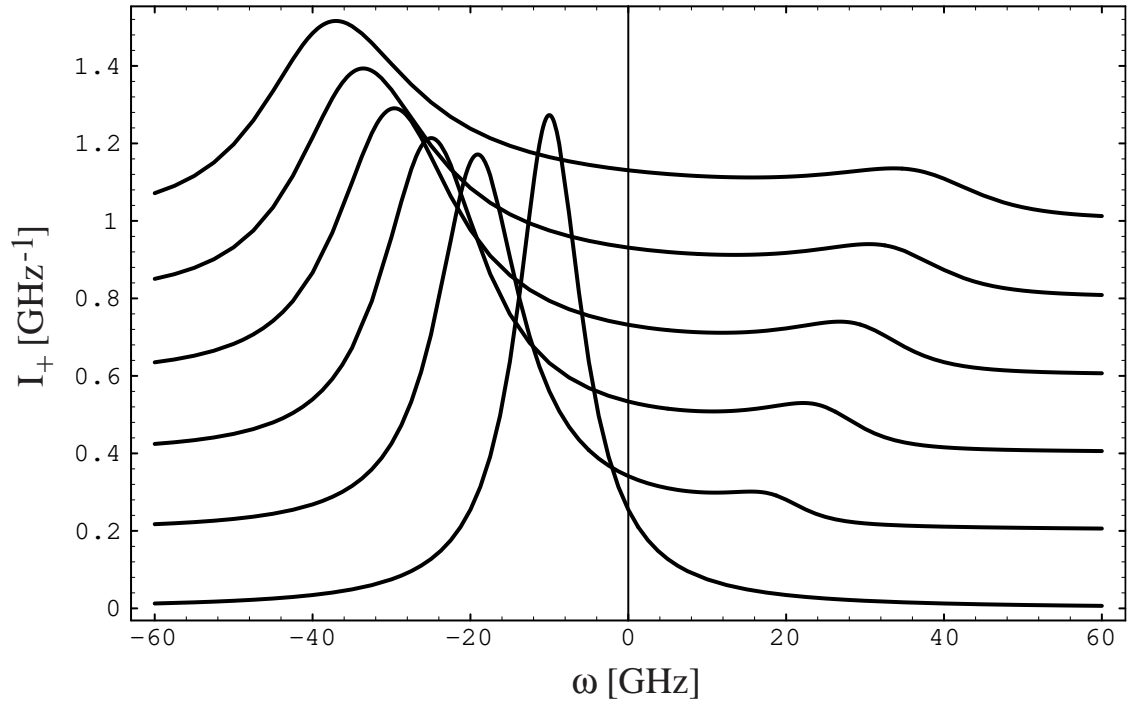
\end{document}